\documentclass[a4paper,twocolumn]{esapub} 
\usepackage{graphicx}
\usepackage{times}

\title{Angular momentum transport in the solar supergranulation layer}
\author{G\"unther R\"udiger}
\author{Manfred K\"uker}
\affil{Astrophysikalisches Institut Potsdam, 14482 Potsdam, Germany}

\def\Om{\it \Omega}
\def\nT{$\nu_{\rm T}$}     

\def\aa{{A\&A}\ }
\def\an{{Astron. Nachr.}\ }

\renewcommand{\vec}[1]{\mbox{\boldmath$#1$}}
\begin{document}

\maketitle

\begin{abstract}
The eddy viscosity in the solar supergranulation layer is derived from the 
observed rotational shear by computing theoretical rotation laws for the 
outermost parts of the solar convection zone using the results from numerical 
simulations of rotating convection as input. By varying the eddy viscosity, the 
results can be tuned to match the observations. The value of $1.5 \times 
10^{13}$ cm$^2$/s found for the eddy viscosity is considerably larger than the 
eddy magnetic diffusivity derived from the sunspot decay. The results are 
checked by comparison of the horizontal cross correlations of the velocity 
fluctuations with the observed Ward profile. 
\end{abstract}
\section{Observations}
\begin{figure}
\includegraphics[width=8cm,height=5cm]{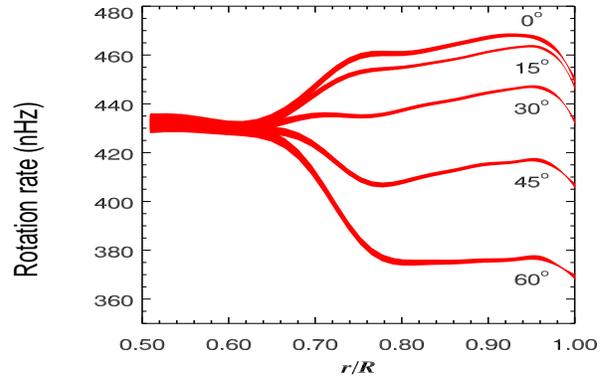}
\caption{The internal solar rotation as found by helioseismology. Note the 
increase of the rotation rate with depth in the outermost layer. The 
maximum lies about 0.05 solar radii beneath the photosphere. (Graph 
courtesy NSF's National Solar Observatory).}
\label{fig1}
\end{figure}

Sunspots rotate about 4\% faster than the solar surface plasma  at all 
latitudes. Helioseismology reveals a corresponding maximum of the angular 
velocity rather close to the surface, as shown in Fig.~\ref{fig1}. Such a clear 
subrotation of the outermost layer of the convection zone is easiest understood 
as the result  of angular momentum conservation of fluid elements with purely 
radial motions. In this domain of the solar convection zone, however, the gas 
motion is predominantly horizontal. Fluctuating fields with dominant 
horizontal intensity should simply produce superrotation rather than the 
observed subrotation (cf. R\"udiger 1989)

From the proper motions of sunspot groups, the faster of which tend to 
move toward the equator, Ward (1965) found the positive value of $\approx 0.1 $ 
(deg / day)$^2 \approx 2 \times 10^7$ m$^2$/s$^2$ for the corresponding 
horizontal cross correlation on the northern hemisphere. More recent 
observations found smaller but always positive values (Nesme-Ribes, Ferreira, 
\& Vince 1993).

%
%
\section{The Reynolds stress}
The mean motion of a turbulent fluid is governed by the Reynolds equation,
\begin{equation} \label{reynolds}
  \rho \left [ \frac{\partial \vec{\bar{u}}}{\partial t}
      + (\vec{\bar{u}}\cdot \nabla)
       \vec{\bar{u}} \right ] =  - \nabla \cdot (\rho Q)
	   - \nabla p + \rho \vec{g},
\end{equation}
where
\begin{equation}
Q_{ij}=\langle u_i(\vec{x},t)' u_j(\vec{x},t)' \rangle
\end{equation}
is the correlation tensor of the velocity fluctuations, $\vec{u}'$. Two 
components of the correlation tensor are relevant for the transport of angular 
momentum:
\begin{eqnarray}
Q_{r\phi} &=& - \nu_{\rm T} r {\partial \Om \over \partial r} \sin\theta + 
I V \sin\theta \label{qrf} \\
Q_{\theta\phi}&=& - \nu_{\rm T} {\partial \Om \over \partial \theta} \sin\theta 
+ I H \cos\theta, \label{qtf} 
\label{3}
\end{eqnarray}
where \nT is the eddy viscosity, $\Om$ the rotation rate, V and H are 
dimensionless functions, and 
\begin{equation}
I= \sqrt{\langle u_r'^2\rangle \langle u_\phi'^2\rangle},
\label{4}
\end{equation}
the turbulence intensity. The eddy viscosity and the turbulence intensity are 
related through
\begin{equation}
\nu_{\rm T} = {\tilde \nu I \over \Om}.
\label{5}
\end{equation}
In Eqs.~\ref{qrf} and \ref{qtf}, the terms containing \nT describe the usual 
eddy viscosity part of the Reynolds stress while the non-diffusive terms 
containing $I$ represent the $\Lambda$-effect, which drives the differential 
rotation.

To compute the rotation pattern, the quantities $\tilde{\nu}$, $I$, $V$, and 
$H$ are needed. While the intensity $I$ can be estimated from the convective 
velocity, as computed by mixing-length theory, $\tilde{\nu}$, $V$, and $H$ have 
to be determined separately. Kitchatinov \& R\"udiger (1993), and Kitchatinov, 
Pipin, \& R\"udiger (1994) derived approximations using the second order 
approximation. Both the eddy viscosity and the $\Lambda$ coefficients are 
functions of the Coriolis number,
\begin{equation}
 \Omega^* = 4 \pi \frac{\tau}{P_{\rm rot}},
\end{equation}
where $\tau$ is the convective turnover time and $P_{\rm rot}$ the rotation 
period. 
For very slow rotation, $\Omega^* \ll 1$, angular momentum is transported 
outwards, and there is no horizontal $\Lambda$ effect. For rapid rotation, the 
horizontal flux of angular momentum becomes dominant. The radial transport 
vanishes in the equatorial region and is negative at the poles. 
\begin{figure}[h]
\centering
 \hspace{-0.5cm}
\includegraphics[height=5.0cm,width=7.5cm]{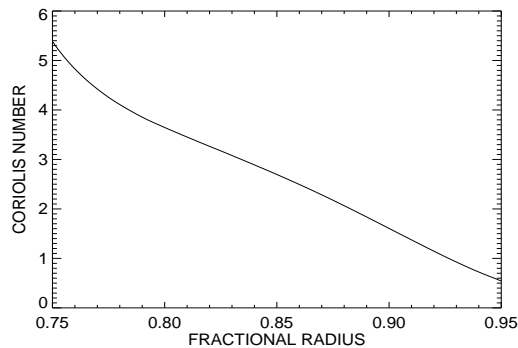}\hfill
\caption{Coriolis number as function of the fractional radius in the bulk of 
the solar convection zone.}
\label{oms}
\end{figure} 

Figure \ref{oms} shows the variation of the Coriolis number with radius in the 
bulk of the solar convection zone, as computed with the model of K\"uker \& 
Stix (2001). The Coriolis number is greater than unity in the bulk of the 
convection zone, but smaller in the granulation and supergranulation layers.
Thus, by comparing the time scales of rotation and convection, we find that the 
bulk of the solar convection zone is rapidly rotating, while the uppermost 
layers are in a state of slow rotation.

\begin{figure}[h]
\centering
\includegraphics[width=7.5cm,height=5cm]{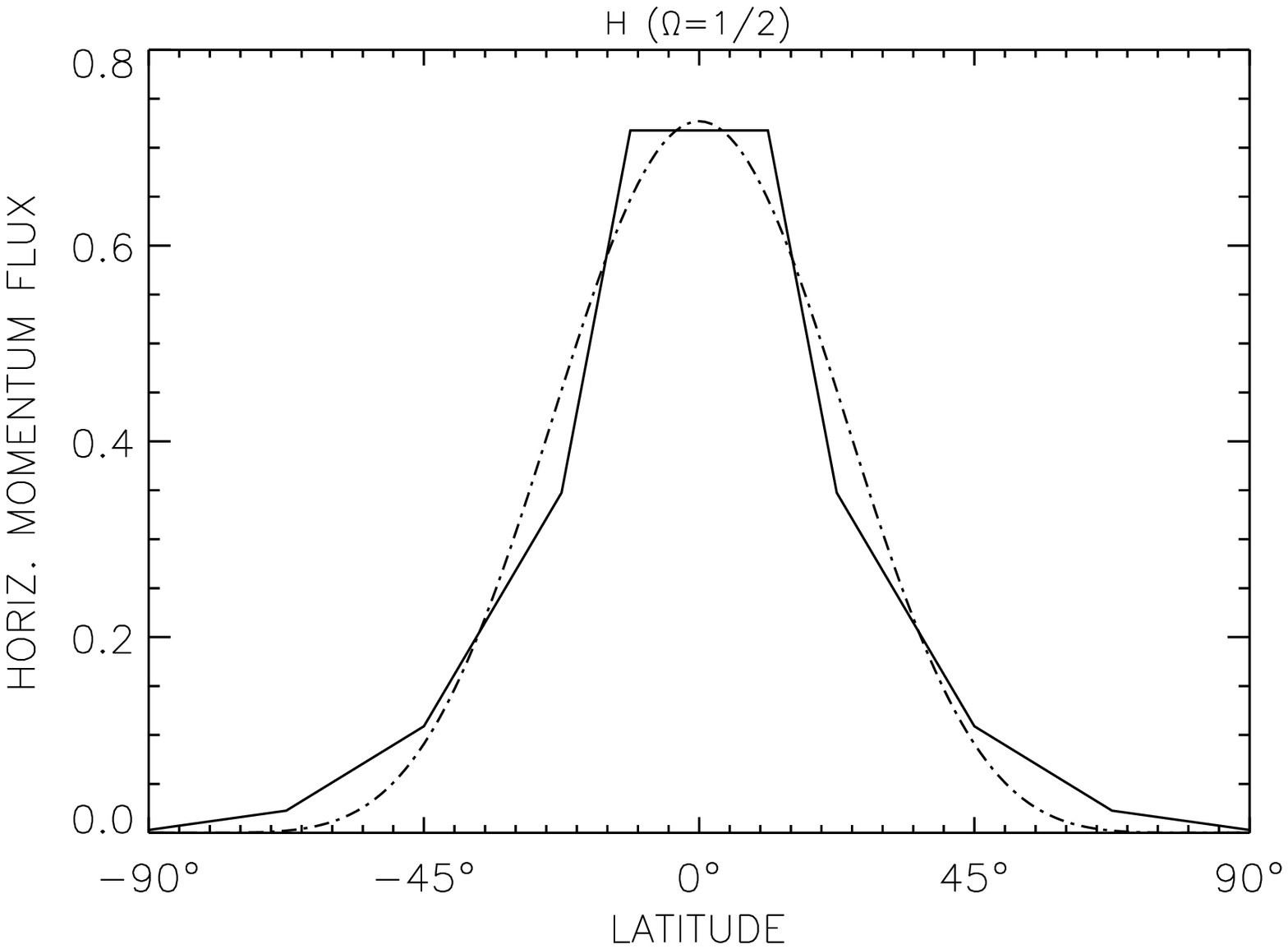}
\includegraphics[width=7.5cm,height=5cm]{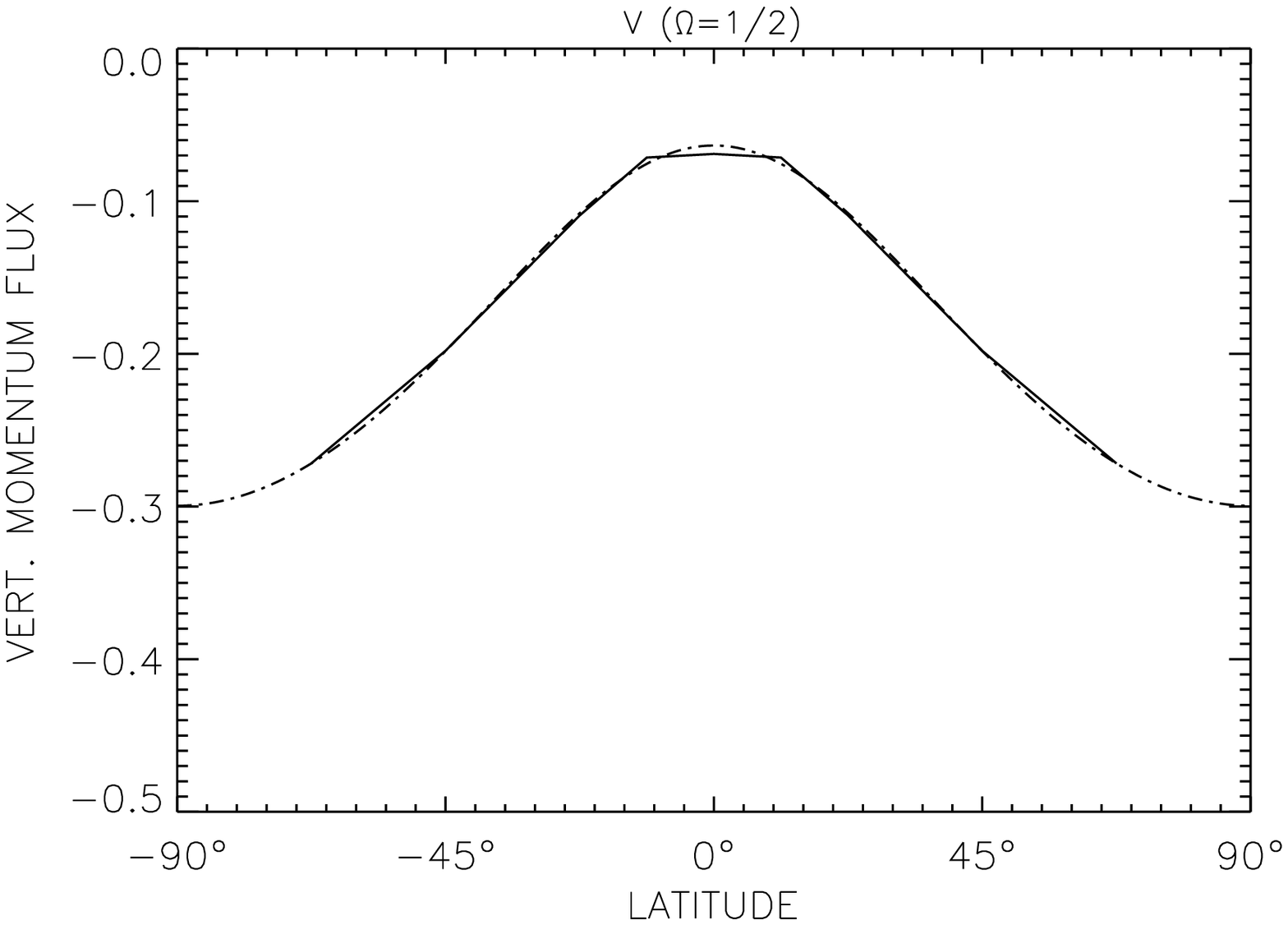}
\caption{The horizontal (top) and vertical (bottom) components of the 
non-diffusive parts of the Reynolds stress, derived from RHD simulations. The 
dash-dotted lines denote the analytical fit used in the computations.}
\label{fig3}
\end{figure}

The rotation laws derived with the expressions from Kitchatinov \& R\"udiger 
(1993), and Kitchatinov, Pipin, \& R\"udiger (1994) for the 
Reynolds stress reproduce the observations remarkably well in the bulk of the 
convection zone, but lack the negative radial shear in the supergranulation 
layer (K\"uker, R\"udiger, \& Kitchatinov 1993, Kitchatinov \& R\"udiger 1995, 
K\"uker \& Stix 2001). We therefore follow a different approach to determine 
the parameters of the stress tensor in the uppermost part of the convection 
zone. Chan (2001) carried out 3D RHD simulations of rotating convection in 
f-planes. Figure \ref{fig3} shows the $V$ and $H$ coefficients as function of 
the colatitude as derived from the simulations. The horizontal component shows 
a strong peak at the equator, while the vertical angular momentum transport is 
strongest at the poles. Contrary to the predictions of Kitchatinov \& R\"udiger 
(1993) for slowly rotating convection, the radial transport of angular momentum 
is directed inwards, and strongest close to the poles rather than at the 
equator. The horizontal transport exceeds the radial, and is strongly 
concentrated at low latitudes.
\section{Results}
We solve the Reynolds equation using a time-explicit finite difference method 
for the outer 5\% of the solar convection zone with the Reynolds stress as 
described above.  As the horizontal shear is mainly generated in 
the bulk of the convection zone which has been excluded from the computations, 
we require 
\begin{equation}
  \Omega(\theta)= \Omega_0 (0.7 + 0.3 \sin^2\theta),
\end{equation}
at the lower boundary. The upper boundary is assumed stress-free, i.e. $Q_{r 
\phi} = 0$. 

In Eqs.~\ref{qrf} and \ref{qtf}, the intensity, $I$ appears in the diffusive as 
well as in the non-diffusive part of the stress tensor. It therefore has no 
influence on the solution, which is stationary, but merely determines the time 
scale on which perturbations are damped. Fixing the average rotation period to 
28 d, the only free parameter is the viscosity parameter $\tilde{\nu}$, which 
is varied between 0 and 1 to get the best possible agreement between the 
computed and the observed rotation laws.

\begin{figure}[h]
\centering
\includegraphics[width=7.0cm,height=5cm]{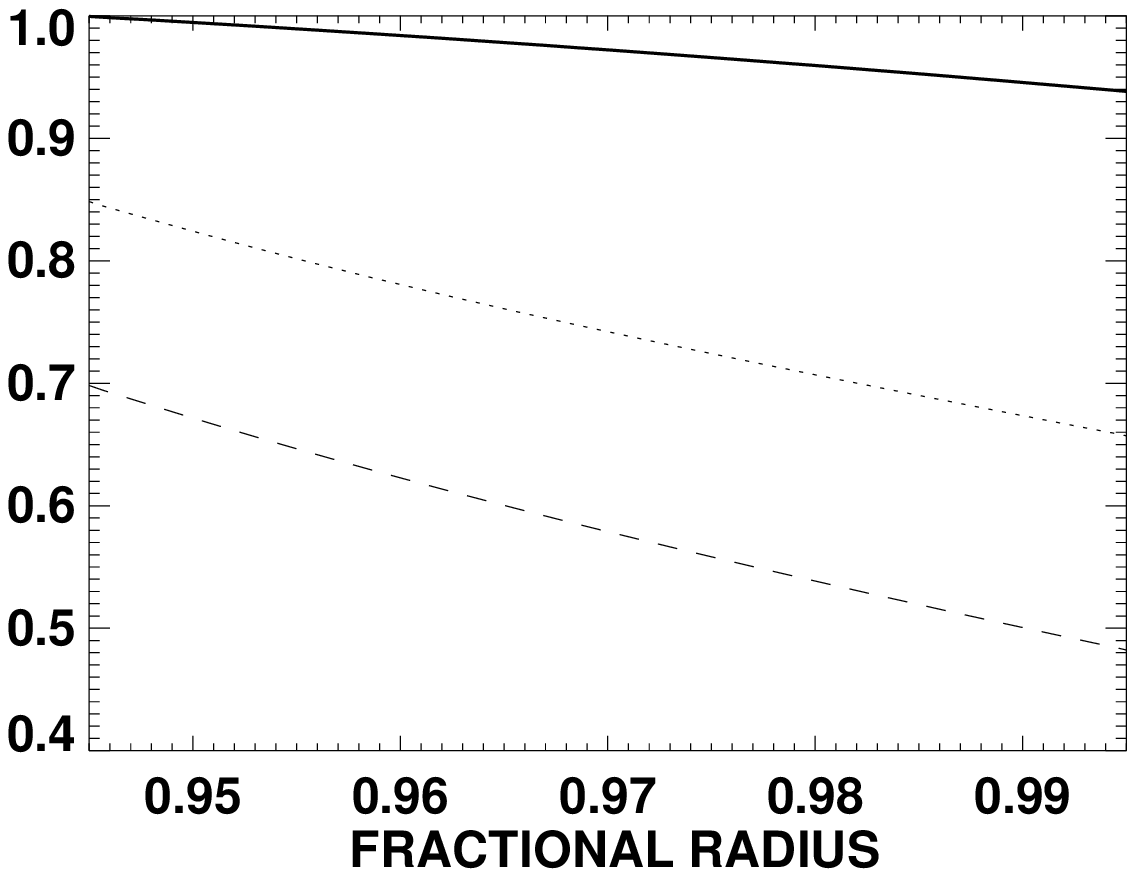}
\includegraphics[width=7.0cm,height=5cm]{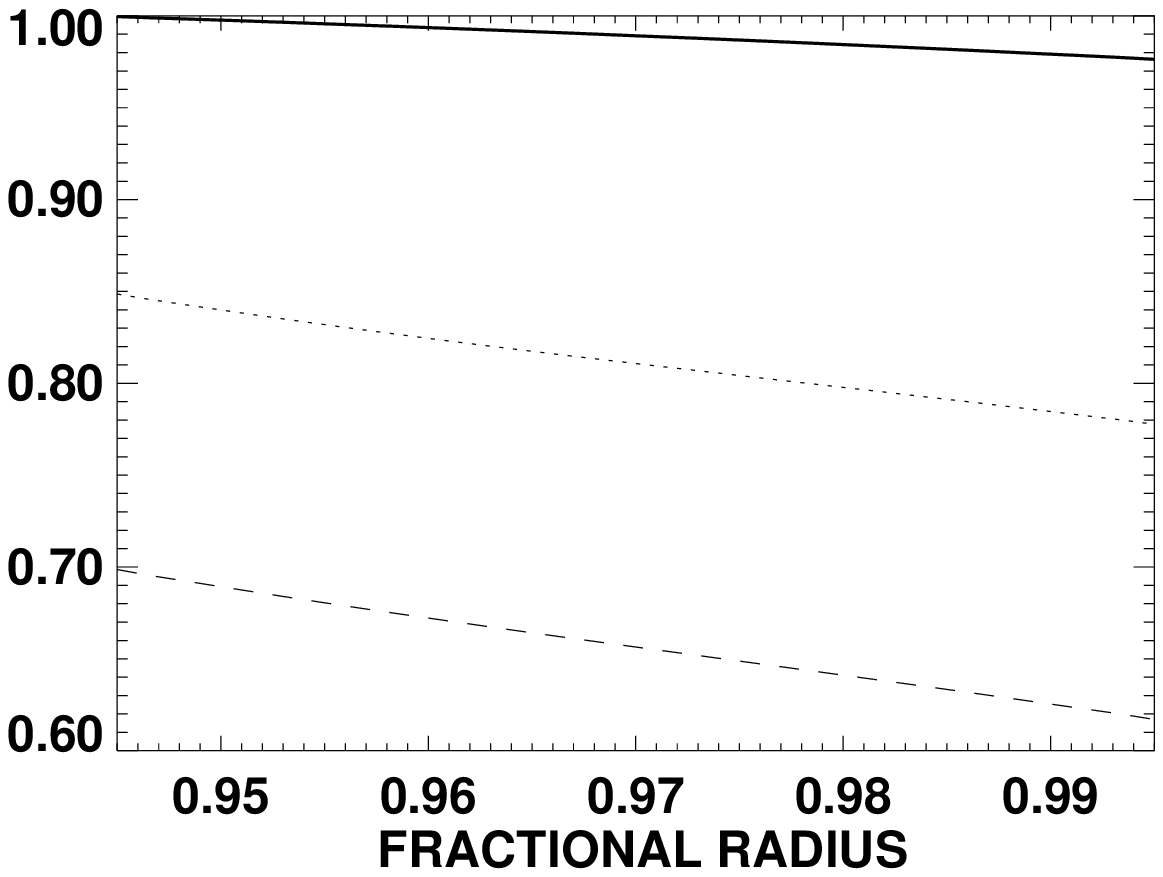}
\includegraphics[width=7.0cm,height=5cm]{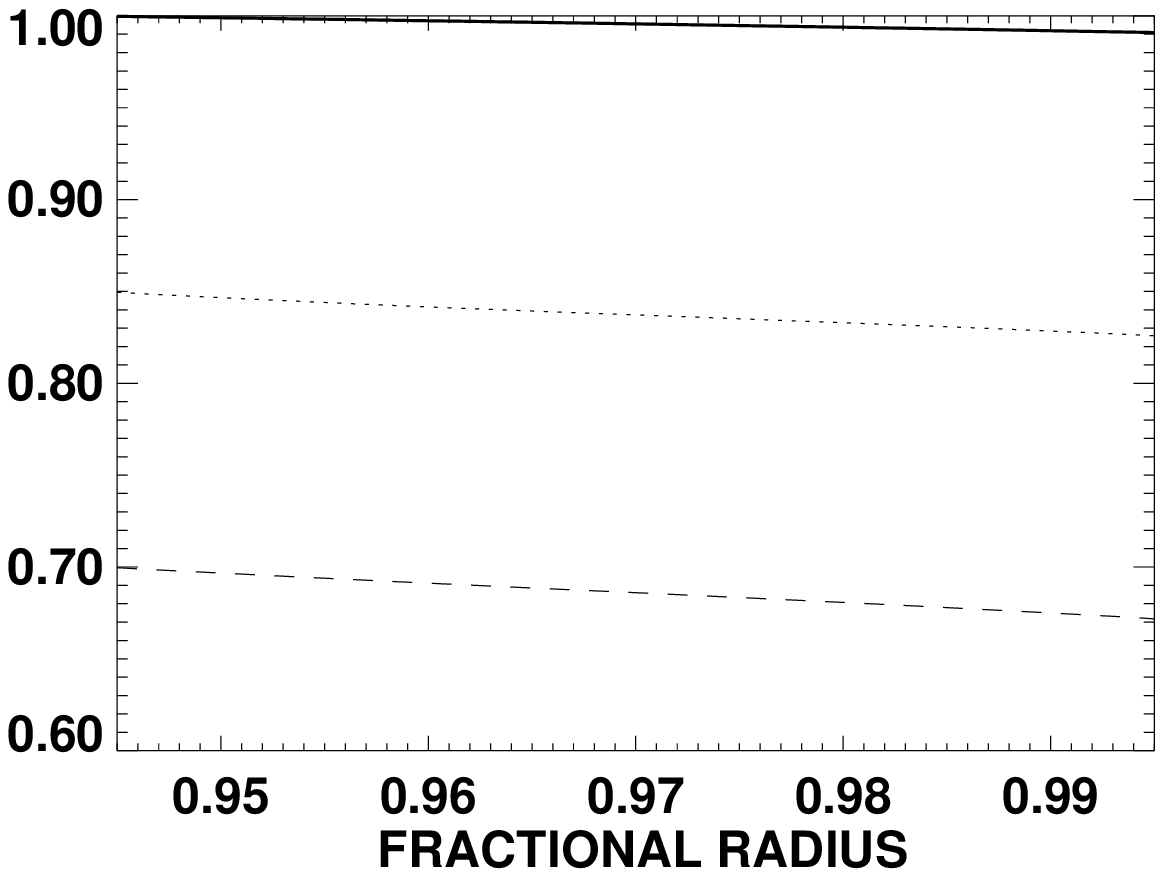}
\caption{The normalized rotation rate at the equator (solid lines), 45 degrees 
latitude (dotted), and the poles (dashed) as a function of the fractional 
radius in the solar supergranulation layer as resulting from theory. From left  
to right: $\tilde{\nu} = 0.03, 0.1, 0.3$.}
\label{fig2}
\end{figure} 

Figure \ref{fig2} shows the rotation pattern for three different values of 
$\tilde{\nu}$. While, due to the lower boundary condition, there are no 
significant differences in the latitudinal shear, the radial shear strongly 
varies. Comparison with the observed profiles in Fig.~\ref{fig1} shows that 
the $\tilde{\nu}=0.1$ case approximately matches the observations, while the 
$\tilde{\nu}=0.03$ case produces too much, and the $\tilde{\nu}=0.3$ case not 
enough shear. We conclude that $\tilde{\nu}=0.1$ is the appropriate choice. 

\begin{figure}[h]
\centering
\includegraphics[height=5.0cm,width=7.5cm]{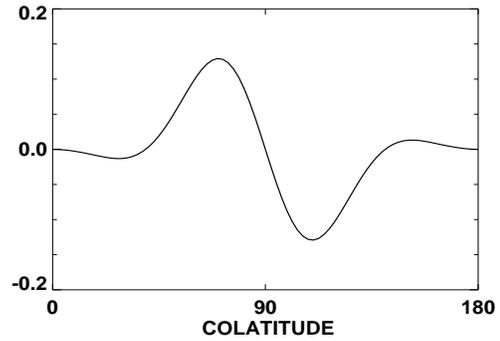}
\caption{Theoretical Ward  profile for $\tilde{\nu}=0.1$.}
\label{fig5}
\end{figure}
   
In Figure \ref{fig5}, the theoretical Ward profile, 
\begin{equation}
W =  - {\tilde \nu \over \Om} {\partial \Om \over 
\partial \theta} \sin\theta + H  \cos\theta,
\label{9.2}
\end{equation}
of the surface rotation law is shown for $\tilde{\nu} =0.1$. At low latitudes, 
the sign is positive in the northern and negative in the southern hemisphere. 
At higher latitudes, the sign is negative in the northern and positive in the 
southern hemisphere. In the equatorial region, the amplitude is roughly the 
same as that observed by Ward (1965), and the signs agree, too. 
%
%

With a typical value of 200 m/s for the convection velocity in the 
supergranulation layer, a rotation rate $\Omega = 2.7 \times 10^{-6} {\rm 
s}^{-1}$, and $\tilde{\nu}=0.1$, we find
\begin{equation}
  \nu_{\rm T} \simeq 1.5 \times 10^{13} {\rm cm}^2/{\rm s}
\end{equation}
for the eddy viscosity. This is two orders of magnitude larger than the value 
of $10^{11}$ cm$^2$/s derived for the magnetic diffusivity from sunspot decay. 
For the ratio between viscosity and magnetic diffusivity, the magnetic Prandtl 
number,
\begin{equation}
  {\rm Pm} = \frac{\nu_{\rm T}}{\eta_{\rm T}}, 
\end{equation}
we thus fund a value as large as 150.


\end{document}